\documentclass[aps,pre,preprint,showpacs,showkeys,onecolumn,superscriptaddress]{revtex4-1}
\usepackage[dvips]{color}
\usepackage{graphicx}
\usepackage{amssymb}
\usepackage{amsmath}
\usepackage[latin1]{inputenc}
\input epsf.sty

\def\sc {\scriptscriptstyle} 
\def\spc {\hskip 0.3mm}

\def\p {\textsf{p}}

\def\Q {{\mathcal Q}}

\def\cp {c^\prime}

\def\wpi {\,\widehat{{\! \varpi}}}

\begin{document}

\title{On the kinetic theory of vehicular traffic flow: Chapman-Enskog expansion  
versus Grad's moment method}


\author{W. Marques Jr.}

\affiliation{
Departamento de F\'{\i}sica, Universidade Federal do Paran\'a, 
81531-990, Curitiba, Brazil}

\author{A. R. M\'endez}
\affiliation{Departamento de Matem\'aticas Aplicadas y Sistemas, 
Universidad Aut\'onoma Metropolitana, 01120 Cuajimalpa, M\'exico}

\begin{abstract}
\noindent Based on a Boltzmann-like traffic equation for aggressive drivers we construct 
in this paper a second-order continuum traffic model 
which is similar to the Navier-Stokes equations for viscous fluids by applying two well-known 
methods of gas-kinetic theory, namely: the Chapman-Enskog method and the method of moments of Grad. 
The viscosity coefficient appearing in our macroscopic traffic model is not introduced in an ad hoc way 
- as in other second-order traffic flow models - but comes into play through the derivation of a 
first-order constitutive relation for the traffic pressure. Numerical simulations show that our 
Navier-Stokes-like traffic model satisfies the anisotropy condition and produces numerical 
results which are consistent with our daily experiences in real traffic.
\end{abstract}


\maketitle

\section{Introduction}

\noindent Because of the great variety of phenomena present in the motion of vehicles along highways or in 
urban roads, traffic dynamics has attracted the attention of 
a large number of researchers in the last decades. Traditionally, 
three different approaches can be used to study traffic flow problems,
namely: a purely microscopic approach in which the acceleration of a 
driver-vehicle unit is determined by other vehicles moving in traffic flow,
a macroscopic approach which describes the collective motion of the
vehicles as the one-dimensional compressible flow of a fluid, and a
mesoscopic approach which specifies the individual behavior of the vehicles by
means of probability distribution functions.

A reading of relevant literature shows that since 1955, when Lighthill and Whitham~\cite{R1} proposed 
the first continuum model to describe traffic flow, much progress has been made in
the development of macroscopic (fluid-type) models on the one hand, and of
microsocopic (follow-the-leader) models on the other hand. The first mesoscopic
(or gas-kinetic) traffic flow model just appeared in 1960, when Prigogine and
Andrews~\cite{R2} wrote a Boltzmann-like equation to describe the time evolution of a
one-vehicle distribution function in a phase-space where the position and
the velocity of the vehicles plays a role. Until the 1990s, mesoscopic traffic
models do not get much attention from scientists due to their lack of ability
to describe traffic operations outside of the free-flow regime. Additionally, 
compared to macroscopic traffic flow models, gas-kinetic traffic models have a
large number of independent variables that 
increase the computational complexity. However, in the last decade, the 
scientific community's interest by mesoscopic traffic models resurrected with
the publication of some works that apply these models to derive macroscopic
traffic models (see, for example, the papers of Helbing~\cite{R3}, Hoogendoorn
and Bovy~\cite{R4}, Wagner et al.~\cite{R5}). Macroscopic equations for relevant traffic
variables can be derived from a Boltzamnn-like traffic equation by averaging
over the instantaneous velocity of the vehicles. This is a well-known procedure
in the kinetic theory, nevertheless its application leads to a closure problem,
i.e., there are some quantities which must be evaluated with constitutive
relations in order to obtain a system of closed equations. The analogy with 
well-established methods of the kinetic theory of gases - such as the
Chapman-Enskog method~\cite{R6} or the method of moments of Grad~\cite{R7} - gives us a clue 
to proceed, provided we have at least a local equilibrium distribution function.

By assuming that motorists drive aggressively~\cite{R8} we apply both the 
Chapman-Enskog method and the method of moments of Grad to construct a second-order 
continuum traffic model which is similar to the Navier-Stokes model for viscous
fluids. In the Chapman-Enskog method, constitutive relations are constructed 
at sucessive levels of approximation by expanding the distribution
function in power of the mean free path, while in Grad's moment method the
distribution function is approximated by an expansion in orthonormal
polynomials where the coefficients are related to the moments of the
distribution function. The dependence of our constitutive relation for the
traffic pressure with the velocity gradient in non-equilibrium situations, 
drive us to define a traffic viscosity coefficient which, in our case, depends on the traffic state through the density 
and the mean velocity of the vehicles. As several second-order continuum 
traffic models, there exists in our Navier-Stokes-like traffic model a characteristic speed 
that is greater than the average flow velocity. The existence of this faster characteristic speed means 
that the motion of the vehicles will be influenced by the traffic conditions behind them. 
This seems to be a drawback of our second-order model since one fundamental principle of traffic flow 
is that vehicles are anisotropic and respond only to frontal stimuli. However,
as pointed out by Helbing and Johansson~\cite{R9} and J. Yi
et al.~\cite{R10}, this faster characteristic
speed does not represent a theoretical inconsistency to our Navier-Stokes
traffic model since it is related to an eigenmode that decays quickly and,
therefore, it cannot emerge by itself. Besides, we check the anisotropic
behavior of our second-order continuum traffic model by performing the
simulations of a traffic situation where a discontinuity is present, namely: the
removal of a blockade scenario.

The organization of the paper is as follows: in Section 2 we present our kinetic
traffic model for aggressive drivers. In Section 3 we construct a second-order continuum traffic model 
- which is similar to the Navier-Stokes equations for viscous fluids - by applying
both the Chapman-Enskog method and the method of moments of Grad. In
Section 4 we present the results of our numerical simulation. Finally, we give
in Section 5 a summary.

\section{Kinetic Traffic Model}

\noindent In a mesoscopic description the state of a vehicle at 
a given instant of time is represented by points in phase space, where the coordinates are 
the position $x$ and the velocity $c$ of the vehicle. Thus, in analogy with the kinetic 
theory of gases, an one-vehicle distribution function $f(x,c,t)$ can be defined in such a 
way that $f(x,c,t)\,dx\,dc$ gives at time $t$ the number of vehicles in the road interval 
between $x$ and $x+dx$ and in the velocity interval between $c$ and $c+dc$. For 
an uni-directional single-lane road without entrances and exits, the one-vehicle 
distribution function satisfies the kinetic traffic equation~\cite{R11}
\begin{gather}
\frac{\partial f}{\partial t}+c\frac{\partial f}{\partial x}
+\frac{\partial}{\partial c}\left(f\spc \frac{dc}{dt}\right)=\Q(f,f),
\end{gather}
where the interaction term
\begin{gather}
\Q(f,f)= \int_c^\infty (1-\p)(\cp-c)\spc f(x,c,t)\spc f(x,\cp,t)\spc d\cp
-\int_0^c (1-\p)(c-\cp)\spc f(x,c,t) f(x,\cp,t)\spc d\cp
\end{gather}
describes the decelaration processes due to slower vehicles which can cannot 
be immediately overtaken. The first part of the interaction term corresponds to
situations where a vehicle with velocity $\cp$ must decelerate to velocity $c$
causing an increase of the one-vehicle distribution function, while the second
one describes the decrease of the one-vehicle distribution function due to
situations in which vehicles with velocity $c$ must decelerate to even slower
velocity $\cp$. The~derivation of the interaction term is based on the
following hypotheses: (i) vehicles are regarded as point-like objects, (ii) the
slowing down process has the probability $(1-\p)$, where $\p$ is the
probability of passing, (iii) the velocity of the slow vehicle is not affected by 
interactions or by being passed, (iv) there is no braking time, (v) only two-vehicle 
interactions are considered and (vi) vehicular chaos is assumed, in such way that the
two-vehicle distribution function can be factorized. The individual acceleration term 
appearing on the left-hand side of the kinetic traffic equation can be modeled by assuming 
that vehicles moving with velocity $c$ accelerate exponentially to their desired velocity 
$c_{\sc 0}=c_{\sc 0} (x,c,t)$ with a relaxation time $\tau$, i.e. 
\begin{gather}
\frac{dc}{dt}=\frac{c_{\sc 0}-c}{\tau}.
\end{gather}
In fact, the desired velocity of the vehicles is determined by the average balance among several traffic 
parameters like legal traffic regulations, weather conditions, road conditions
and drivers personality, i.e., it is a phenomenological function. Despite of
the variety of traffic parameters that determines the desired velocity of the
vehicles, we shall consider in this work the simple relation~\cite{R8}
\begin{gather}
c_{\sc 0}=w c,
\end{gather}
where $w=w(x,t)>1$ is a positive parameter that depends on traffic conditions 
along the road. On driver's level, relation (4) indicates that the desired velocity of the
vehicles increases as their velocity increases, which is 
a common feature of aggressive drivers. By assuming that the aggressiveness
parameter $w$ depends on position and time only through 
the vehicular density, we shall consider the following constitutive relation
\begin{gather}
w=1+\left(w_{\sc c}-1\right)
\frac{\rho}{\rho_{\sc c}}\left(\frac{\rho_{\sc 0}-\rho}{\rho_{\sc 0}-\rho_{\sc c}}\right)^{\dfrac{\rho_{\sc 0}-\rho_{\sc c}}{\rho_{\sc c}}}
\end{gather}
where $\rho_{\sc 0}$ is the jam (or maximum) density, $\rho_{\sc c}$ is the
critical vehicle density (i.e. the density value which maximizes the aggressiveness
parameter) and $w_{\sc c}=w(\rho_{\sc c})$ is the maximal value for drivers'
aggressiveness. Expression (5) tell us that: (i) when traffic is very dilute
the aggressiveness parameter must tend to one, since in this limit vehicles
move along the highway at its desired
speed, (ii) as the vehicle density increases, the
velocity of the vehicles starts to decrease, and drivers try to compensate this
situation by increasing their aggressiveness, (iii) when the density exceeds its
critical value, it becomes increasingly difficult for the vehicles to move at its desired
speed and, in this case, the drivers have no choice except to reduce its
aggressiveness, (iv) when the highway is fully occupied, the vehicles can no
longer move and the aggressiveness parameter must tend to one. Figure 1 shows the dependence of the aggressiveness
parameter on the vehicular density for $\rho_{\sc c}/\rho_{\sc 0}=0.2$ and
$w_{\sc c}=1.2$. Lastly, it is important to mention that the constitutive relation
(5) is not a unique model to the aggressiveness parameter, but just a sound one. 

\begin{figure}[t]
\centering
\includegraphics[width=0.70\textwidth]{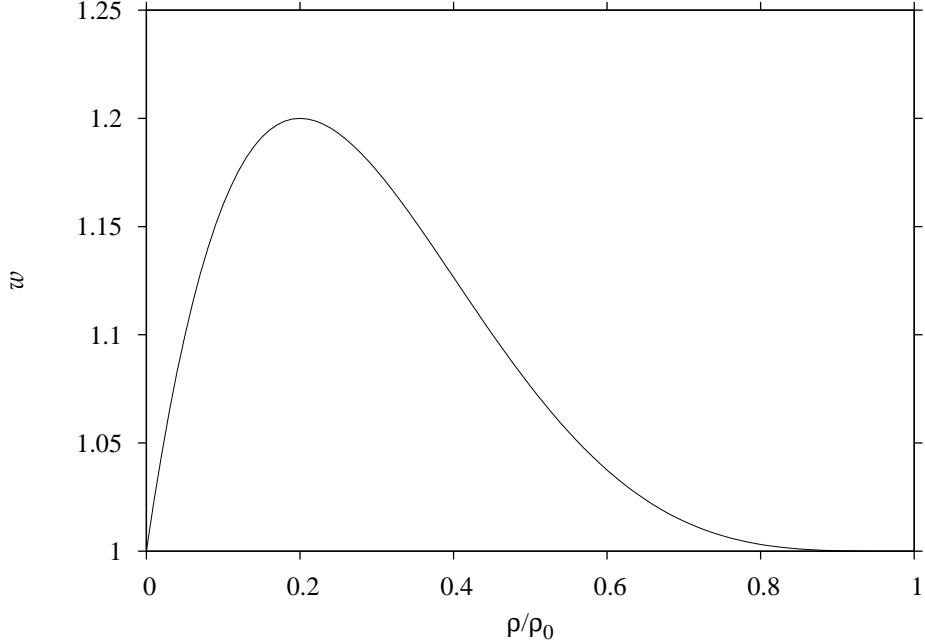}
\caption{Dependence of the aggressiveness parameter on the vehicular density}
\end{figure}

\section{Second-Order Continuum Model}

\noindent The kinetic traffic equation (1) allows the derivation of balance
equations for macroscopic traffic variables like the vehicular density
\begin{gather}
\rho (x,t)=\int_0^\infty f(x,c,t)\spc dc
\end{gather}
and the average velocity
\begin{gather}
v(x,t)=\int_0^\infty c\spc \frac{f(x,c,t)}{\rho (x,t)}\spc dc.
\end{gather}
The integration of the kinetic traffic equation over all values of the actual 
velocity of the vehicles yields the continuity equation 
\begin{gather}
\frac{\partial \rho}{\partial t}+\frac{\partial \rho v}{\partial x}=0,
\end{gather}
while the traffic momentum equation
\begin{gather}
\frac{\partial \rho v}{\partial t}+\frac{\partial}{\partial x}\left(\rho
  v^2+\varpi\right)=\rho\spc \frac{w-1}{\tau}\spc v-\rho\spc (1-\p)\spc \varpi
\end{gather}
follows through the multiplication of the kinetic traffic equation with $c$
and the integration over all values of the actual velocity of the vehicles. In the derivation 
of the traffic momentum equation we have introduced the traffic pressure 
\begin{gather}
\varpi (x,t)=\int_0^\infty (c-v)^2 f(x,c,t)\spc dc, 
\end{gather}
and used relations (3) and (4). At this point it is important to emphasize that the balance equations (8) and
(9) can only be obtained if the one-vehicle distribution function satisfies the following boundary
conditions:
\begin{gather}
\lim_{c\rightarrow 0} f(x,c,t)=0\quad\quad \hbox{and}\quad\quad 
\lim_{c\rightarrow \infty} f(x,c,t)=0.
\end{gather}
Based on the continuity and momentum equations, we can construct a
second-order continuum traffic model by specifying the traffic pressure in
terms of the vehicular density, the average velocity and their spatial
gradients. Since there are a variety of possible constitutive relations which
can be borrowed from fluid dynamics, we shall restrict ourselves here to the
derivation of a constitutive relation for the traffic pressure which is similar
to the usual Navier-Stokes relation for ordinary viscous fluids, i.e., a constitutive relation
written in terms of the density, the average velocity and their first-order
spatial gradients. One can achieve this goal by applying, for example, the
Chapman-Enskog method or the method of moments of Grad, as they are developed in the
kinetic theory of gases. In the Chapman-Enskog
method, constitutive relations are constructed at sucessive levels of
approximation by expanding the distribution function in power of the mean free
path, while in Grad's moment method the gas-kinetic equation is replaced by a set
of balance equations for the moments of the distribution function. To close 
this set of equations, the distribution function is approximated by an
expansion in orthonormal polynomials, where the coefficients are related to the
moments of the distribution function. Then, by applying an iteration procedure
in the resulting system of field equations, it is possible to derive
first-order constitutive relations. Here, we shall apply both
methods to derive a first-order constitutive relation for the traffic pressure
which is similar to the Navier-Stokes relation for viscous fluids.

\subsection{Chapman-Enskog Method}

\noindent The basic idea of the Chapman-Enskog method is to expand 
the distribution function into a power series in the rarefaction parameter $\varepsilon$ as 
\begin{gather}
f=f^{\sc (0)}+\frac{f^{\sc (1)}}{\varepsilon}+\frac{f^{\sc (2)}}{\varepsilon^2}+\dots
=\sum_{r=0}^\infty \frac{f^{\sc (r)}}{\varepsilon^r}
\end{gather}
where $f^{\sc (r)}$ represents successive approximations to the distribution
function. In traffic flow problems, the rarefaction parameter is a
dimensionless number defined as 
\begin{gather}
\varepsilon=\frac{\text{L}}{\lambda}
\end{gather}
where $\text{L}$ is the road length and $\lambda$ denotes 
the mean distance traveled by a vehicle between successive interactions. 
Based on the value of the rarefaction paramater, it is possible to identify two traffic flow
regimes: (i) the rarefied regime, when $\varepsilon \ll 1$,
 and  (ii) the continuum regime, when $\varepsilon \gg 1$. In the rarefied regime, 
the processes of acceleration and decelaration of the
vehicles can be neglected and the general solution of the kinetic traffic
equation in this regime reads
\begin{gather}
f(x,c,t)=f(x-ct,c,t=0). 
\end{gather}
Otherwise, in the continuum regime, the processes of
acceleration and deceleration of the vehicles can not be neglected, so that 
the distribution function is determined by both the individual acceleration law and the interaction
term. To determine the distribution function in this limit, it is appropriate
to write the kinetic traffic equation (1) in the following dimensionless form 
\begin{gather}
\frac{\partial f}{\partial t}+c\frac{\partial f}{\partial x}
+\varepsilon\,\frac{\partial}{\partial c}\left(f\,\frac{dc}{dt}\right)
=\varepsilon\int_0^\infty (1-\p)\left(\cp-c\right)f(x,c,t) f(x,\cp,t) d\cp
\end{gather}
where the position is given in units of the road length, the velocity is given in
units of the average velocity, the time is given units of the mean flow time
(the mean time required for vehicles to traverse the road) and the vehicular
acceleration is given in units of $v^2/\lambda$. 

Integral equations for the distribution function approximations can be easily 
obtained if we insert the expansion (12) into the dimensionless kinetic traffic
equation (15) and equate equal power of the scale parameter. Hence, we obtain 
\begin{gather}
\frac{\partial}{\partial c}
\left(f^{\sc (0)}\spc \frac{dc}{dt}\right)=\Q (f^{\sc (0)},f^{\sc (0)})
\end{gather}
and
\begin{gather}
\frac{\partial f^{\sc (r-1)}}{\partial t}+c\frac{\partial f^{\sc (r-1)}}{\partial x}
+\frac{\partial}{\partial c}
\left(f^{\sc (r)}\spc \frac{dc}{dt}\right)
=\Q (f^{\sc (0)},f^{\sc (r)})
\qquad (r\ge 1)
\end{gather}
where
\begin{gather}
\Q(f^{\sc (0)},f^{\sc (r)})=\sum_{n=0}^r\int_0^\infty (1-\p)\left(\cp-c\right) f^{\sc (n)}
f^{\prime \sc (r-n)}\, d\cp.
\end{gather}
By assuming that the vehicular density and the mean velocity are only
determined by the zeroth-order approximation of the distribution function, the
solution of the integral equation (16) with the individual acceleration law (3)
leads to
\begin{gather}
f^{\sc (0)}=\frac{\alpha}{\Gamma (\alpha)}\frac{\rho}{v}\left(\frac{\alpha c}{v}\right)^{\!\alpha-1}
\exp\left(-\frac{\alpha c}{v}\right)
\end{gather}
where
\begin{gather}
\alpha=\frac{\rho\left(1-\p\right)v\spc\tau}{w-1}.
\end{gather}
Expression (19) for the zeroth-order distribution function tell us that the
velocity of the vehicles is gamma-distributed with a shape parameter $\alpha$
and a rate parameter $\beta=v/\alpha$ when drivers drive agressively on the
highway. To gain an insight into the shape parameter, let us calculate the
velocity variance (or velocity dispersion) in the zeroth-order approximation,
i.e., 
\begin{gather}
{\mathit \Theta}=\frac{\displaystyle \int_0^\infty (c-v)^2 f^{\sc (0)} dc}{\displaystyle \int_0^\infty f^{\sc (0)} dc}
=\frac{v^2}{\alpha}.
\end{gather}
From the above expression we verify that the velocity variance depends
quadratically on the mean velocity, a fact which can be used to identify the
inverse of the shape parameter as the prefactor of the velocity variance. The
experimental data reported by Shvetsov and Helbing~\cite{R12} demonstrate that the
prefactor of the velocity variance is almost constant at the low-density
region, otherwise it can be taken as a function of the vehicular density. In
this work, we shall take this prefactor or, rather, the shape parameter as a
constant whose value is greater than unity, so that our gas-kinetic-like
traffic model is restricted to low densities.

In order to determine the first-order approximation to the one-vehicle distribution
function we introduce expression (19) into the integral equation  
\begin{gather}
\frac{\partial f^{\sc (0)}}{\partial t}+c\frac{\partial f^{\sc (0)}}{\partial x}
+\frac{\partial}{\partial c}
\left(f^{\sc (1)}\spc \frac{dc}{dt}\right)
=\Q (f^{\sc (0)},f^{\sc (1)})
\end{gather}
which follows from (17) by taking $r=1$. By utilizing the constraints
\begin{gather}
\int_0^\infty f^{\sc (r)}\spc dc=0\quad\quad \text{and}\quad\quad 
\int_0^\infty c\spc f^{\sc (r)}\spc dc=0
\end{gather}
which are valid for $r\geq 1$ we obtain
\begin{gather}
f^{\sc (0)}\biggl[\frac{{\dot \rho}}{\rho}
+\frac{v}{\rho}\Bigl(\frac{c}{v}-1\Bigr)\frac{\partial\rho}{\partial x}
+\alpha \Bigl(\frac{c}{v}-1\Bigr)\frac{{\dot v}}{v}
+\alpha \Bigl(\frac{c}{v}-1\Bigr)^{\!2}\frac{\partial v}{\partial x}
\biggr]+\frac{\partial}{\partial c}
\biggl(f^{\sc (1)}\spc \frac{dc}{dt}\biggr)
=-\rho (1-\p)(c-v)f^{\sc (1)}
\end{gather}
where the dot denotes the material time derivative. Integration of equation
(24) over all values of the actual velocity leads to the continuity equation
\begin{gather}
\frac{{\dot \rho}}{\rho}=-\frac{\partial v}{\partial x}.
\end{gather}
Besides, if we multiply equation (24) with $c$ and integrate over all values of
the actual velocity we get 
\begin{gather}
\alpha\spc \frac{{\dot v}}{v}=-\frac{v}{\rho}\spc \frac{\partial\rho}{\partial x}
-2\spc \frac{\partial v}{\partial x}
-\alpha \left(1-\p\right) v \int_0^\infty \Bigl(\frac{\cp}{v}-1\Bigr)^{\!2}\spc f^{\prime \sc (1)}d\cp.
\end{gather}
If equations (25) and (26) are used to eliminate the material time derivatives
appearing in equation~(24) we obtain 
\begin{gather}
f^{\sc (0)}\biggl[\alpha \Bigl(\frac{c}{v}-1\Bigr)^{\!2}-2\spc
  \Bigl(\frac{c}{v}-1\Bigr)-1\biggr]\frac{\partial v}{\partial x}
+\frac{\partial}{\partial c}
\biggl(f^{\sc (1)}\spc \frac{w-1}{\tau}\spc c\biggr)
+\alpha \Bigl(\frac{c}{v}-1\Bigr)\frac{w-1}{\tau}\spc f^{\sc
  (1)}\nonumber\\[3mm]
=f^{\sc (0)}\alpha \left(1-\p\right) v\spc \Bigl(\frac{c}{v}-1\Bigr)
\int_0^\infty \Bigl(\frac{\cp}{v}-1\Bigr)^{\!2}\spc f^{\prime \sc (1)}d\cp.
\end{gather}
Based on equation (27) we shall write the first-order approximation to
one-vehicle distribution function as
\begin{gather}
f^{\sc (1)}=f^{\sc (0)}\left[\spc a_{\sc 0}+a_{\sc 1}\spc \Bigl(\frac{c}{v}-1\Bigr)+a_{\sc
    2}\spc \Bigl(\frac{c}{v}-1\Bigr)^{\!2}\spc\right]\frac{\partial v}{\partial x}
\end{gather}
where $a_{\sc 0}$, $a_{\sc 1}$ and $a_{\sc 2}$ are position and time-dependent
coefficients to be determined. Insertion of expression (28) into
constraints (23) and equation (27) leads to 
\begin{gather}
a_{\sc 0}=\frac{a_{\sc 1}}{2}=-\frac{a_{\sc 2}}{\alpha}=\frac{\tau}{2\left(w-1\right)}.
\end{gather}
Note that the dependence of the expansion coefficients with position and time comes 
through the parameter $w$ that measures the average agressiveness of drivers. Hence, the 
first-order approximation to the distribution function reads
\begin{gather}
f^{\sc (1)}=-\frac{f^{\sc (0)}}{2}\spc \frac{\tau}{(w-1)}
\biggl[\alpha \Bigl(\frac{c}{v}-1\Bigr)^{\!2}-2\spc 
  \Bigl(\frac{c}{v}-1\Bigr)-1\biggr]\frac{\partial v}{\partial x}.
\end{gather}

Once the one-vehicle distribution function is a known function of the basic
traffic variables and their first-order spatial gradients, the traffic pressure
can be directly calculated as
\begin{gather}
\varpi=\int_0^\infty (c-v)^2\left(f^{\sc (0)}+f^{\sc (1)}\right) dc=
\frac{\rho v^2}{\alpha}
-2\, \frac{\rho v^2}{\alpha}\spc \tau_{\sc 0}\spc
\Bigl(\frac{\alpha+1}{\alpha}\Bigr)\spc
\frac{\partial v}{\partial x}
\end{gather}
where
\begin{gather}
\tau_{\sc 0}=\frac{\tau}{2\left(w-1\right)}.
\end{gather}
The constitutive relation (31) for the traffic pressure has a similar form to
the Navier-Stokes relation for viscous fluids since in non-equilibrium
situations both constitutive relations depend on the velocity gradient. Based
on this similarity, it is possible to define a traffic viscosity coefficient
\begin{gather}
\mu=\mu (\rho,v)=2\, \frac{\rho v^2}{\alpha}\spc \tau_{\sc 0}\spc
\Bigl(\frac{\alpha+1}{\alpha}\Bigr)
\end{gather}
which depends on the traffic state through the vehicular density and the
average velocity. At this point, it is important to mention that a similar
constitutive relation for the traffic pressure was derived by Velasco and
Marques Jr.~\cite{R8} by applying a simplified version of the Chapman-Enskog method to
the reduced Paveri-Fontana traffic equation~\cite{R13}. In their formalism, the collective
relaxation time $\tau_{\sc 0}$ appears as a free adjustable parameter of order
of the mean vehicular interaction time, and it was introduced by means of a
relaxation approximation performed in the interaction term.

\subsection{Grad's Moment Method}

\noindent In Grad's moment method a macroscopic description of traffic flow is
based on macroscopic traffic variables like the vehicular density, the average 
velocity and the central moments of the distribution function
\begin{gather}
m_k(x,t)=\int_0^\infty (c-v)^k f(x,c,t)\spc dc\quad\quad (k\ge 2).
\end{gather}
The balance equations governing the dynamical behavior of these macroscopic
traffic variables are the continuity equation (8), the traffic momentum equation (9)
and the balance equations
\begin{gather}
\frac{\partial m_k}{\partial t}+\frac{\partial}{\partial x}\left(m_k v+m_{k+1}\right)
+k\spc m_k\frac{\partial v}{\partial x}-k\spc \frac{m_{k-1}}{\rho}\frac{\partial 
\varpi}{\partial x}-k\spc\frac{w-1}{\tau}\spc m_k\nonumber\\[3mm]
=-\rho\spc (1-\p)\biggl(m_{k+1}-k\spc \frac{m_{k-1}}{\rho}\spc \varpi\biggr).
\end{gather}
In deriving the balance equations (35) for the central moments we have
multiplied the kinetic traffic equation (1) by $(c-v)^k$, and subsequently
integrated with respect to the actual velocity of the vehicles. 

Clearly, we can see that the balance equations (8), (9) and (35) form a non-closed system of field equations
for the determination of the moments $\rho$, $v$ and $m_k$, 
since the balance equation for the central moment $m_k$ contains the central 
moment $m_{k+1}$ which is not a priori related to the lower order moments. 
The dependence of the central moment $m_{k+1}$ upon the moments $\rho$, $v$ and
$m_k$ can be attained if we know the distribution function as a function of $\rho$, $v$ and
$m_k$. In the method of the moments such normal solution is found by an expansion
around a zeroth-order distribution function~(19), i.e., we write the distribution function as  
\begin{gather}
f(x,c,t)=f^{\sc (0)}(x,c,t)\sum_{n=0}^\infty C_n(x,t) P_n (c)
\end{gather}
where $C_n(x,t)$ are position and time-dependent expansion coefficients and $P_n (c)$ are
orthonormal polynomials in the actual velocity of the vehicles. Since in the
zeroth-order approximation the velocity of the vehicles is gamma-distributed,
it is possible to construct the orthonormal polynomials $P_n(c)$ by applying
the condition
\begin{gather}
\int_0^\infty {\mathit \Phi}(s) P_n(s) P_m(s)\,ds=\delta_{nm}
\end{gather}
where $s=\alpha c/v$ is the dimensionless instantaneous velocity and 
\begin{gather}
{\mathit \Phi} (s)=\frac{s^{\alpha-1}e^{-s}}{\Gamma (\alpha)}
\end{gather}
is the probability density function of the gamma distribution. From the
orthonormality condition (37) we verify that the first polynomials read
\begin{gather}
P_0(s)=1,\\[2mm] 
P_1(s)=\frac{s-\alpha}{\sqrt{\alpha}},\\[2mm] 
P_2(s)=\frac{s^2-2\spc (\alpha+1)\spc s+\alpha\spc (\alpha+1)}{\sqrt{2\spc \alpha\spc (\alpha+1)}},\\[2mm]
P_3(s)=\frac{s^3-3\spc (\alpha+2)\spc s^2+3\spc (\alpha+1)(\alpha+2)\spc s
-\alpha\spc (\alpha+1)(\alpha+2)}{\sqrt{6\spc \alpha\spc (\alpha+1)(\alpha+2)}}.
\end{gather}
We can easily verify from the above expressions that the orthonormal polynomials $P_n(s)$
are related to the associated Laguerre polynomials (see the textbook of Arfken~\cite{R14}) and they are given by
the formula
\begin{gather}
P_n (s)=\frac{(-1)^n}{s^{\alpha-1}e^{-s}}\sqrt{ \frac{\Gamma (\alpha)}{n!\, 
\Gamma (\alpha+n)}}\spc \frac{d^n}{ds^n}\left(s^{n+\alpha-1} e^{-s}\right).
\end{gather}
By using the orthonormality condition (37) the
position and time-dependent expansion coefficients~$C_n$ can be determined as follows:
\begin{gather}
\int_0^\infty P_n(c) f(x,c,t)\,dc=\rho \sum_{m=0}^\infty C_m \int_0^\infty {\mathit \Phi} (s) P_n(s) P_m(s) \,
ds=\rho C_n.
\end{gather}
Thus, we conclude that the expansion
coefficients $C_n$ are related directly to the moments of the distribution
function. For example, the first coefficients read  
\begin{gather}
C_{\sc 0}=1,\\[3mm]
C_{\sc 1}=0,\\[3mm]
C_{\sc 2}=\sqrt{\frac{\alpha}{2\spc (\alpha+1)}}\spc \frac{\varpi-\varpi_{\sc 0}}{\varpi_{\sc 0}},\\[3mm]
C_{\sc 3}=\sqrt{\frac{2\spc \alpha}{3\spc (\alpha+1)(\alpha+2)}}\spc \biggl(
\frac{\phi-\phi_{\sc 0}}{\phi_{\sc 0}}-3\spc \frac{\varpi-\varpi_{\sc 0}}{\varpi_{\sc 0}}\biggr),
\end{gather}
where
\begin{gather}
\phi(x,t)=m_3(x,t)=\int_0^\infty (c-v)^3 f(x,c,t)\spc dc
\end{gather}
is the third-order central moment. Besides,
\begin{gather}
\varpi_{\sc 0}(x,t)=\int_0^\infty \!\! (c-v)^2 f^{\sc (0)}(x,c,t)\spc dc
=\frac{\rho v^2}{\alpha}
\end{gather}
and
\begin{gather}
\phi_{\sc 0}(x,t)=\int_0^\infty \!\! (c-v)^3 f^{\sc (0)}(x,c,t)\spc dc
=2\spc \frac{\rho v^3}{\alpha^2}
\end{gather}
are the values of the second and third-order central moments in the zeroth-order approximation. 
Insertion of the position and time-dependent coefficients into the expansion 
of the distribution function allows us to write it in terms of the velocity polynomials and the 
macroscopic traffic variables. Note that each coefficient in the expansion 
of the distribution function introduces a new macroscopic traffic variable, 
so that it is possible to choose which relevant variables we want to use in 
our macroscopic traffic description. 

Let us now construct a continuum traffic flow model based only on three traffic
variables, namely: the vehicular density, the average velocity and the traffic
pressure. In this case, the balance equations governing the dynamical behavior
of traffic variables are the continuity 
equation (8), the momentum equation (9) and traffic pressure equation 
\begin{gather}
\frac{\partial \varpi}{\partial t}+\frac{\partial}{\partial x}\left(\varpi v+\phi\right)
+2\spc\varpi \frac{\partial v}{\partial x}-2\spc \frac{w-1}{\tau}\spc\varpi=-\rho\spc (1-\p)\spc\phi.
\end{gather}
The balance equations (8), (9) and (52) becomes a system of
field equations for the determination of $\rho$, $v$ and $\varpi$, if a
relationship can be established between these variables and the third-order
central moment $\phi$. In order to achieve this
goal, expansion (36) for the distribution function is taken with $C_n(x,t)=0$ for
$n\ge 3$, so that we have  
\begin{gather}
f=f^{\sc (0)}\left\{1+\frac{s^2-2\spc (\alpha+1)\spc s+\alpha\spc
    (\alpha+1)}{2\spc (\alpha+1)}\spc \frac{\varpi-\varpi_{\sc 0}}{\varpi_{\sc 0}}\right\}.
\end{gather}
A comparison of the distribution function for the first order Chapmann-Enskog
expansion (30) and the one obtained by Grad's method (53) shows that the last
expression (i) involves an extended set of variables, note that in this
formalism the traffic pressure is an independent variable, and (ii) depends only on
the macroscopic variables not on the gradients. If the traffic pressure is
calculated assuming (53), we realize that the result is different from (31). 
To reconcile these differences we will carry out an standard iterative
procedure known in the kinetic theory of gases as the Maxwellian iteration procedure.

Insertion of the distribution function (53) into expression (49)
leads, after a simple integration, to the following constitutive relation for 
the third-order central moment:
\begin{gather}
\phi=3\spc\frac{\phi_{\sc 0}}{\varpi_{\sc 0}}\left(\varpi-\frac{2}{3}\spc \varpi_{\sc 0}\right).
\end{gather}
If we introduce the constitutive relation (54) into the balance equations (8),
(9) and (52) we obtain a system of field equations for $\rho$, $v$ and $\varpi$ or, 
equivalently, for $\rho$, $v$ and $\wpi$, where $\wpi=\varpi-\varpi_{\sc 0}$ is
the so-called traffic pressure deviator. Hence, after some algebra, we get 
\begin{gather}
\frac{\partial\rho}{\partial t}+v\spc \frac{\partial\rho}{\partial x}
+\rho\spc \frac{\partial v}{\partial x}=0,\\[3mm]
\rho\spc \frac{\partial v}{\partial t}+\frac{\varpi_{\sc 0}}{\rho}\spc \frac{\partial \rho}{\partial x}
+(\alpha+2)\spc \frac{\varpi_{\sc 0}}{v}\spc \frac{\partial v}{\partial x}
+\frac{\partial \wpi}{\partial x}=
\rho\spc \frac{w-1}{\tau}\spc v 
-\rho\, (1-\p)\spc (\varpi_{\sc 0}+\wpi),\\[3mm]
\frac{\partial \wpi}{\partial t}
+\Bigl(\frac{\alpha+4}{2}\Bigr)
\spc\frac{\phi_{\sc 0}}{\varpi_{\sc 0}}\spc \frac{\partial \wpi}{\partial x}
+3\spc \Bigl(\frac{\alpha+2}{\alpha}\Bigr)\spc\wpi \spc \frac{\partial v}{\partial x} 
+2\spc\varpi_{\sc 0} \Bigl(\frac{\alpha+1}{\alpha}\Bigr)\spc \frac{\partial v}{\partial x}
=-\frac{\wpi}{\tau_{\sc 0}}.
\end{gather}
As in the kinetic theory of gases, one can transform the balance equation (57) 
into an approximate constitutive relation for the traffic pressure deviator by applying a method akin with the 
Maxwellian iteration procedure~\cite{R15}. For the first-iteration step we insert, on the
left-hand side, the value of the traffic pressure deviator in the zeroth-order
approximation, namely $\wpi=0$, and get, on the right-hand side, the first iterated value
\begin{gather}
\wpi=-2\spc\frac{\rho v^2}{\alpha}\spc \tau_{\sc 0}\spc
\Bigl(\frac{\alpha+1}{\alpha}\Bigr)\spc
\frac{\partial v}{\partial x}.
\end{gather}
The above expression for the traffic pressure deviator is identical
to that one obtained by the Chapman-Enskog procedure, a fact that 
allow us to affirm that Chapman-Enskog expansion and Grad's moment method 
are both physically and mathematically equivalent, at least in first-order approximation.

\subsection{Navier-Stokes-like Traffic Model}

\noindent Insertion of the constitutive relation (31) for the traffic pressure
into the balance equations (8) and~(9) leads to the so-called
Navier-Stokes-like traffic model which can be written in the following
conservative form
\begin{gather}
\frac{\partial {\bf U}}{\partial t}
+\frac{\partial {\bf F}({\bf U})}{\partial x}={\bf S}({\bf U})
\end{gather}
where
\begin{gather}
{\bf U}=\begin{pmatrix}
\rho \\[3mm] \rho v
\end{pmatrix},
\quad
{\bf F}({\bf U})=\begin{pmatrix}
\rho v\\[3mm]
\rho v^2+\rho c^2
\end{pmatrix}
\quad\text{and}\quad
{\bf S}({\bf U})=\begin{pmatrix}
0 \\[3mm]
\rho\spc \dfrac{u(\rho,v)-v}{\tau}-b\spc v_x+(\mu v_x)_x
\end{pmatrix}.
\end{gather}
In the above expressions we have introduced the traffic sound speed
$c(v)=\sqrt{\partial \varpi_{\sc 0}/\partial \rho\,}$, the optimal velocity function 
$u(\rho,v)=wv-\tau\spc (1-\p)\spc \varpi_{\sc 0}$ and the anticipation coefficient
\begin{gather}
b\spc (\rho,v)=-\left(\frac{\alpha-1}{2}\right)\spc \frac{\partial \varpi_{\sc 0}}{\partial v}<0.
\end{gather}
In contrast to others macroscopic traffic models, we see that our optimal velocity function 
does not depend only on the vehicular density, but also on the average velocity and that such 
dependence is explicitly determined by the average desired velocity of the vehicles reduced 
by a term arising from deceleration processes due to vehicle interactions. Besides, in our 
macroscopic traffic flow model, traffic viscosity is not introduced in an ad
hoc way - as in some of the most popular second-order contiunuum traffic models - but it 
comes into play via an interation procedure and reflects the way drivers anticipate traffic situation 
on the basis of second-order spatial changes in the average velocity.

Finally, we close this section by remarking that the eigenvalues $\Lambda$ 
of the conservative flux-Jacobian matrix
\begin{gather}
{\bf A}({\bf U})=\frac{\partial {\bf F}({\bf U})}{\partial {\bf U}}=
\begin{pmatrix}
0 & & 1\\[3mm]
-\dfrac{\alpha+1}{\alpha}\spc v^2  & & 2\spc \dfrac{\alpha+1}{\alpha}\spc v
\end{pmatrix}
\end{gather}
determine how traffic information (such as slown-downs and speed-ups) is transmitted in a 
traffic stream. These eigenvalues, also known as characteristic speeds, are found by setting 
\begin{gather}
\text{det}\spc \vert \spc {\bf A}({\bf U})- \Lambda {\bf I} \spc \vert = 0,
\end{gather}
where ${\bf I}$ is the identity matrix. Hence, our Navier-Stokes-like traffic model 
has two real and distinct characteristic speeds, namely:
\begin{gather}
\Lambda_{\sc 1,2}=\frac{\alpha+1\mp \sqrt{\alpha+1}}{\alpha}\spc v.
\end{gather}
Note that one of the characteristic speeds is larger than the average traffic flow velocity
indicating that traffic disturbances propagate in the downstream direction. This
fact has been criticized by Daganzo~\cite{R16} as a major deficiency of most second-order
traffic models. However, as shown by Helbing and Johanson~\cite{R9} in a recent publication, 
this fact does not constitute a theoretical inconsistency of our second-order macroscopic 
traffic model since the perturbation that travels faster than traffic decays
quickly. 

\section{Numerical Example}

\noindent Since the Navier-Stokes-like traffic equations (59) form a hyperbolic
system of partial differential equations which involve smooth as well as
discontinuous solutions, it is suitable to use a conservative numerical
scheme to solve them accurately. However, as pointed out in the literature, 
conservative numerical schemes are only appropriate for solving the homogeneous
part of the system, i.e. when~the source term ${\bf S}({\bf U})$ vanishes. In order to deal with the source
term, the following fractional-step scheme~\cite{R17}
\begin{gather}
{\bf U}_{\text i}^\ast={\bf U}_{\text i}^{\text n}-\frac{\Delta t}{\Delta x}\left( {\bf
    F}_{\text i+1/2}-{\bf F}_{\text i-1/2}\right)\\[3mm]
{\bf U}_{\text i}^{\text n+1}= {\bf U}_{\text i}^\ast+\frac{\Delta t}{2}\spc 
\left({\bf S}({\bf U}_{\text i}^{\text n})+{\bf S}({\bf U}_{\text i}^\ast)\right)
\end{gather}
can be employed, where $\Delta x$ is the grid size, $\Delta t$ is the time
increment and ${\bf F}_{\text i+1/2}={\bf F}({\bf U}_{\text i}^{\text n},
{\bf U}_{\text i+1}^{\text n})$ is the intercell numerical flux. The intercell 
numerical flux  ${\bf F}_{\text i+1/2}$ approximates the time-integral average
of flux across the cell interface between cells $\text i$ and $\text i+1$, 
where the i-th cell is given by the spatial interval between $x_{\text i-1/2}$ 
and $x_{\text i+1/2}$. One can find in the literature several classes of
methods that provide stable and consistent approximations to the intercell numerical
flux. Here, we determine the numerical flux by employing 
a first order upwind-type scheme based on Roe's flux-difference splitting~\cite{R18}.

The application of Roe's scheme to the homogeneous version of our macroscopic traffic
model yields a conservative method whose numerical flux function is computed as
\begin{gather}
{\bf F}_{\text i+1/2}=\frac{{\bf F}({\bf U}_{\text i})
+{\bf F}({\bf U}_{\text i+1})}{2}-\frac{1}{2}\sum_{\text{k}} \sigma_{\sc
\text{k}}\spc \vert \Lambda_{\sc \text{k}}\vert \spc {\bf e}_{\sc \text{k}},
\end{gather}
where $\Lambda_{\sc \text{k}}$, $\sigma_{\sc \text{k}}$ and ${\bf e}_{\sc
  \text{k}}$ are respectively the eigenvalues, the wave strengths and the
right-eigenvectors of the flux-Jacobian matrix ${\bf A}({\bar {\bf U}})$
at Roe's average state ${\bar {\bf U}}$. For the Navier-Stokes-like traffic
model these quantities are expressed as:
\begin{gather}
\Lambda_{\sc 1,2}=\frac{\alpha+1\mp \sqrt{\alpha+1}}{\alpha}\spc {\bar v},\\[3mm]
\sigma_{\sc 1,2}=\mp \frac{\alpha}{2\,\sqrt{\alpha+1}}\left[ 
\frac{\Delta (\rho v)}{{\bar v}}-\frac{\alpha+1\pm
  \sqrt{\alpha+1}}{\alpha}\,\Delta \rho\right],\\[3mm]
{\bf e}_{\sc 1,2}=\begin{pmatrix}
1\\[3mm] \dfrac{\alpha+1\mp \sqrt{\alpha+1}}{\alpha}\spc {\bar v}
\end{pmatrix},
\end{gather}
where $\Delta {\bf U}={\bf U}_{\text{i+1}}-{\bf U}_{\text i}$. As usual, Roe's average state is determined by
imposing the shock-capturing property
\begin{gather}
\Delta {\bf F}={\bf F}({\bf U}_{\text i+1})-{\bf F}({\bf U}_{\text i})={\bf
  A}({\bar {\bf U}})\,\Delta {\bf U}.
\end{gather}
From (71) we verify that the first equation is identically satisfied, while the
second one gives the following average for the velocity:
\begin{gather}
{\bar v}=\frac{\sqrt{\rho_{\spc\text i}}\, v_{\spc\text i}+\sqrt{\rho_{\spc\text i+1}}\, v_{\spc\text i+1}}
{\sqrt{\rho_{\spc\text i}}+\sqrt{\rho_{\spc\text i+1}}}.
\end{gather}
Note that no representation for average
traffic density is required in our Navier-Stokes-like traffic model, since
Roe's average (72) for the flow velocity completely defines the flux-Jacobian
matrix. Concerning now the discretization of the source term we take
\begin{gather}
{\bf S}({\bf U}_{\text i})=\rho_{\spc \text i}\frac{u(\rho_{\spc \text i},v_{\spc \text i} )-v_{\spc
    \text i}}{\tau}-\frac{b (\rho_{\spc \text i},v_{\spc \text i} )}{2}\spc \frac{v_{\spc \text i+1}-v_{\spc \text i-1}}{\Delta x}
+\frac{\mu_{\spc \text i+1/2}\spc (v_{\spc \text i+1}-v_{\spc \text i})- 
\mu_{\spc \text i-1/2}\spc (v_{\spc \text i}-v_{\spc \text i-1})}{(\Delta x)^2},
\end{gather}
where
\begin{gather}
\mu_{\spc \text i\pm 1/2}=\frac{\mu (\rho_{\spc \text i},v_{\spc \text i})
+\mu (\rho_{\spc \text i\pm 1},v_{\spc \text i\pm 1})}{2}.
\end{gather}

\begin{figure}[t]
\centering
\includegraphics[angle=-90,width=0.75\textwidth]{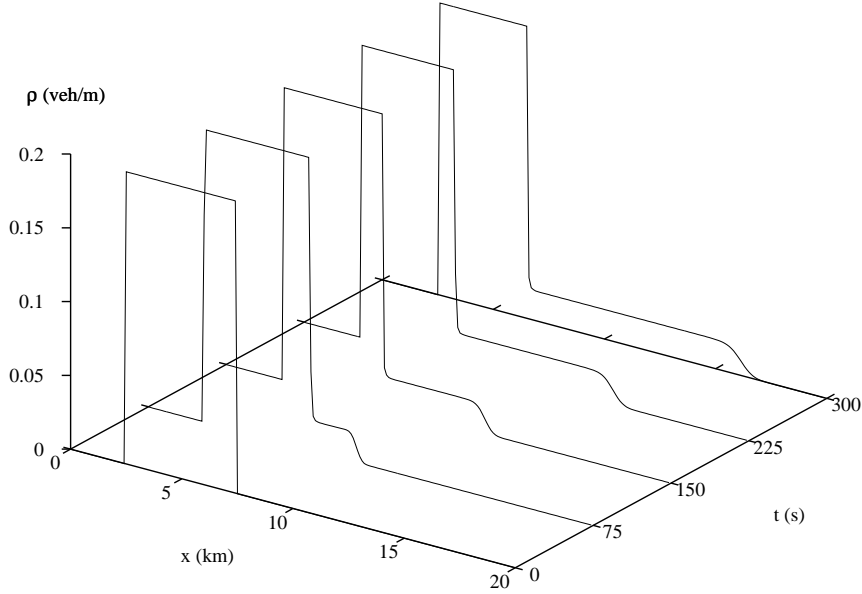}
\vskip 0.3truecm
\caption{Time evolution of the vehicle density for the removal of blockade scenario}
\end{figure}

\noindent Completing the description of the numerical scheme it is worth to mention that
the numerical stability of explicit difference schemes is ensured by the 
Courant-Friedrich-Lewy condition
\begin{gather}
\frac{\Delta t}{\Delta x}\,\text{max}_{_{_{_{_{\!\!\!\!\!\!\!\!\text{k}}}}}} \;\;\vert \Lambda_{\text k} \vert \leq 1
\end{gather}
which guarantees that information propagates only through a single cell at each
time step.

We illustrate now the numerical scheme presented above by considering a
traffic situation where a queue of nearly motionless vehicles is present in a certain road region. At
the initial time, the blockade at the head of the queue is removed and vehicles
flow into the empty part of the roadway. For simulation of this traffic
scenario, we consider the following initial conditions on a 20 km circular
road:
\begin{gather}
\rho(x,0)=\begin{cases}
\,\spc \rho_{\ast}\,, & \hbox{if}\,\,\, 2.5\spc\spc \text{km} < x <
7.5\spc\spc \text{km}\\[3mm]
\,\,\,0\,\,, & \text{elsewhere}
\end{cases}\quad\quad\text{and}\quad\quad
v(x,0)=v_e(\rho(x,0)),
\end{gather}
where $\rho_\ast$ is the vehicle density in the queue and $v_e(\rho)$ is the
equilibrium speed-density function. Here, we assume an equilibrium
speed-density relation of the form (see Bando et al.~\cite{R19})
\begin{gather}
v_e(\rho)=\frac{v_{\sc 0}}{2}\left(\text{tanh}\left(\rho_{\sc 0}/\rho-a\right)
+\text{tanh}\left(a\right)\right),
\end{gather}
where $v_{\sc 0}$ is the free-flow velocity and $a$ is a positive dimensionless
constant. Furthermore, we impose periodic
boundary conditions and choose the following values for the model parameters:
\begin{gather}
\alpha = 125,\quad v_{\sc 0} = 30\,\text{m/s},\quad \rho_{\sc 0}=0.2\,
\text{veh/m},\quad \rho_\ast=0.198\, \text{veh/m},\
 \quad a=3.9\quad \text{and}\quad \tau=8\, \text{s}.
\end{gather}
Regarding the probability of passing it is usual to assume that this quantity depends only on the 
vehicular density in a linear way. However, as pointed out by Hoogendoorn and Bovy~\cite{R20} in their report, an expression for 
the probability of passing that depends both on the vehicular density and the mean velocity can be obtained 
if we set $u(\rho,v)=v_e(\rho)$. By equating the optimal velocity to the equilibrium speed-density function we 
are in fact replacing the individual (microscopic) process of deceleration by some collective (macroscopic) 
relaxation process to an equilibrium traffic state. 

Figures 2 and 3 show that after the removal of the blockade the vehicles at the head
of the queue move downstream into the empty region with the free-flow velocity, while
vehicles at the tail of the queue remain at their location. Although the
traffic conditions upstream are free-flow we observe from these figures that
vehicles do not flow backwards into the empty region, a fact that  allow us to
say that our Navier-Stokes-like traffic model satisfies the anisotropy
condition and produces numerical results which are similar to traffic
operations in real-life traffic. Similar results were obtained by Hoogendoorn~\cite{R21} through numerical
simulations performed on the single user-class version of his macroscopic multiple
user-class traffic flow model. Hoogendoorn's macroscopic multiple user-class
traffic model was derived from mesoscopic principles which encompass
contributions of drivers acceleration towards their user-class specific desired
velocity and contributions resulting from interactions between vehicles of the
same and different classes. Besides, the velocity variance is introduced as an
additional basic field describing deviations from the average velocity within
the user-classes.

\begin{figure}[t]
\centering
\includegraphics[angle=-90,width=0.75\textwidth]{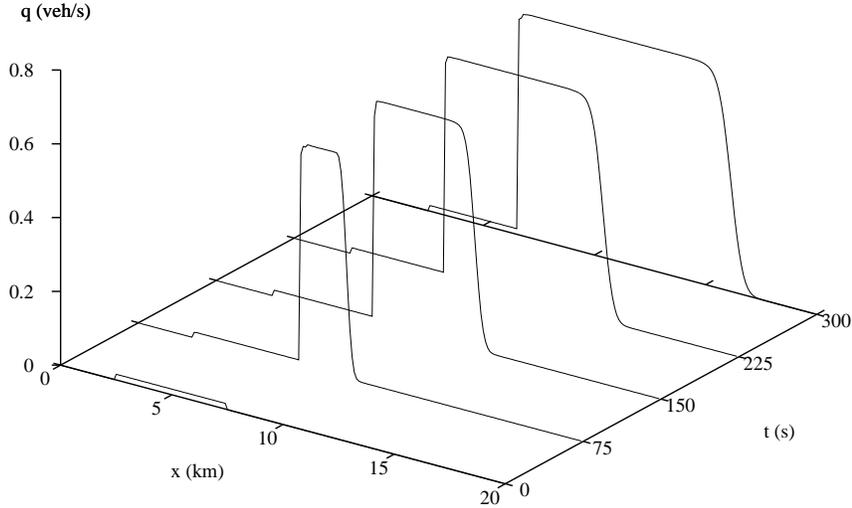}
\vskip 0.3truecm
\caption{Time evolution of the traffic flow for the removal of blockade scenario}
\end{figure}

\section{Summary}

\noindent By appling both Chapman-Enskog expansion and Grad's moment method we constructed a second-order
continuum traffic model which is very similar to the Navier-Stokes model for
viscous fluids. The constitutive relation for the traffic pressure obtained by
the method of moments of Grad is identical to that one obtained by the
Chapman-Enskog procedure, a fact that allow us to affirm that Chapman-Enskog expansion and Grad's moment method 
are both physically and mathematically equivalent, at least in first-order
approximation.  Besides, in contrast to others second-order macroscoscopic traffic
models, our traffic viscosity coefficient - which depends on the traffic state
through the vehicle density and the mean velocity - is not introduced in an ad hoc way,
but comes into play through the derivation of a constitutive relation to the
traffic pressure. Numerical simulation of a traffic scenario where a
discontinuity is present show that our macroscopic traffic model
satisfies the anisotropy condition and produces numerical results which are similar to traffic
operations in real-life traffic.

\end{document}